\documentclass[aps,prl,reprint,superscriptaddress,showpacs]{revtex4-1}

\usepackage[utf8]{inputenc}
\usepackage[dvips]{graphicx}
\usepackage{epsfig}
\usepackage{multirow}
\usepackage{arydshln}
\usepackage{nicefrac}
\usepackage{textcomp}
\usepackage{gensymb}
\usepackage{booktabs}
\usepackage{url}
\usepackage{threeparttable}
\usepackage{amsmath}

\usepackage{color}

\newcommand{\beginsupplement}{%
        \setcounter{table}{0}
        \renewcommand{\thetable}{S\arabic{table}}%
        \setcounter{figure}{0}
        \renewcommand{\thefigure}{S\arabic{figure}}%
     }

\begin{document}

\title{Direct evidence of weakly dispersed and strongly anharmonic optical phonons in hybrid perovskites}

\author{A.C. Ferreira}
\affiliation{Universit\'e Paris-Saclay, CNRS, CEA, Laboratoire L{\'e}on Brillouin,  91191, Gif-sur-Yvette, France}
\affiliation{Univ Rennes, INSA Rennes, CNRS, Institut FOTON - UMR 6082, F-35000 Rennes, France}
\author{S. Paofai}
\affiliation{Univ Rennes, ENSCR, INSA Rennes, CNRS, ISCR - UMR 6226, F-35000 Rennes, France}
\author{A. L\'etoublon}
\affiliation{Univ Rennes, INSA Rennes, CNRS, Institut FOTON - UMR 6082, F-35000 Rennes, France}
\author{J. Ollivier}
\affiliation{Institut laue Langevin, 71 avenue des martyrs, 38000 Grenoble, France}
\author{S. Raymond}
\affiliation{Univ. Grenoble Alpes, CEA, IRIG, MEM, MDN, F-38000 Grenoble}
\author{B.~Hehlen}
\affiliation{Laboratoire Charles Coulomb, UMR 5221 CNRS-Universit\'e de Montpellier, Montpellier, FR-34095, France}
\author{B. Ruffl\'e}
\affiliation{Laboratoire Charles Coulomb, UMR 5221 CNRS-Universit\'e de Montpellier, Montpellier, FR-34095, France}
\author{S. Cordier}
\affiliation{Univ Rennes, ENSCR, INSA Rennes, CNRS, ISCR - UMR 6226, F-35000 Rennes, France}
\author{C. Katan}
\affiliation{Univ Rennes, ENSCR, INSA Rennes, CNRS, ISCR - UMR 6226, F-35000 Rennes, France}
\author{J. Even}\email{Corresponding author:  jacky.even@insa-rennes.fr}
\affiliation{Univ Rennes, INSA Rennes, CNRS, Institut FOTON - UMR 6082, F-35000 Rennes, France}
\author{P. Bourges}\email{Corresponding author: philippe.bourges@cea.fr}
\affiliation{Universit\'e Paris-Saclay, CNRS, CEA, Laboratoire L{\'e}on Brillouin,  91191, Gif-sur-Yvette, France}
\date{\today}

\begin{abstract}
\noindent 

Hybrid organolead perovskites (HOP) have started to establish themselves in the field of photovoltaics, mainly due to their great optoelectronic properties and steadily improving solar cell efficiency. Study of the lattice dynamics is key in understanding the electron-phonon interactions at play, responsible for such properties. Here, we investigate, via neutron and Raman spectroscopies, the optical phonon spectrum of four different HOP single crystals: MAPbBr$_3$, FAPbBr$_3$, MAPbI$_3$, and $\alpha$-FAPbI$_3$. Low temperature spectra reveal weakly dispersive optical phonons, at energies as low as 2-5~meV, which seem to be the origin of the limit of the charge carriers mobilities in these materials. The temperature dependence of our neutron spectra shows as well a significant anharmonic behaviour, resulting in optical phonon overdamping at temperatures as low as 80~K, questionning the validity of the quasi-particle picture for the low energy optical modes at room temperature where the solar cells operate.

\end{abstract}
\maketitle
%
\noindent \textbf{Introduction}\\

Over the last few years, halide perovskites have emerged as a promising class of materials for high-performing photovoltaic (PV) cells~\cite{Correa-Baena2017, Snaith2018, Jena2019}. The hybrid organolead perovskites (HOP) adopt an APbX$_3$ structure, where A is an organic cation (methylammonium, MA, or formamidinium, FA) and X is an halide (Cl, Br, I) and their optoelectronic properties, as well as easy and cost-effective production (from abundant chemical elements), make HOPs not only attractive for PV applications~\cite{Yang2015, Anaraki2016, Tan2017, Kojima2009, Extance2019}, but also for light-emitting devices (LEDs) and many other application based on thin films or even single crystals~\cite{Yuan2017, Akkerman2018, Smith2019, Fu2019}.\\
\indent  Several of the properties responsible for the outstanding performance of hybrid perovskites are connected with electron-phonon interactions, which have been under intense debate \cite{Wright2016, Chen2016, Even2016, Miyata2017, Katan2018}. Besides governing their emission line broadening, phonon scattering is among the factors setting a fundamental intrinsic limit to the mobility of charge carriers in these materials. From the inspection of photoluminescence (PL) lineshape broadening in the high-temperature regime of 3D perovskite thin films, it was deduced that carriers scattering is dominated by Fröhlich coupling between charge carriers and longitudinal optical (LO) phonon modes,  rather than with acoustic phonons \cite{Wright2016, Even2016}. This was later reinforced from the analysis of elastic constants among different 3D hybrid perovskites \cite{Letoublon2016,Ferreira2018}. More, optical phonons were shown to play a central role in the slow carriers relaxation in colloidal quantum dots (CQD)  of FAPbI$_{3}$ and FAPbBr$_{3}$,  from the exciton bright triplet to the dark singlet through a second order process \cite{Fu2018}, with strong consequences for CQD brightness and quantum efficiency of CQD light emitting devices \cite{Fu2018, Tamarat2019}. This mechanism is an alternative to the Rashba effect, which was proposed to lead to an inversion of dark and bright exciton states in CQDs \cite{Becker2018}. Further, the potential of perovskite CQDs for hot-carriers solar cell applications has also been stressed \cite{Li2019}: the suppression of the LO relaxation process to longitudinal acoustic (LA) phonons was attributed to an optical phonon bottleneck effect \cite{Yang2015a}, and later on related to the anharmonicity of the acoustic modes \cite{Yang2017}.  Then, a direct measurement of optical phonons branches with the same methodology as for the acoustic modes\cite{Letoublon2016,Ferreira2018} becomes a necessary step to completely uncover carrier-phonon coupling dynamics and to assess the fundamental intrinsic limit of the mobility of charge carriers in these materials.\\
\begin{figure}[t]
\centering
\includegraphics[width=0.95\linewidth]{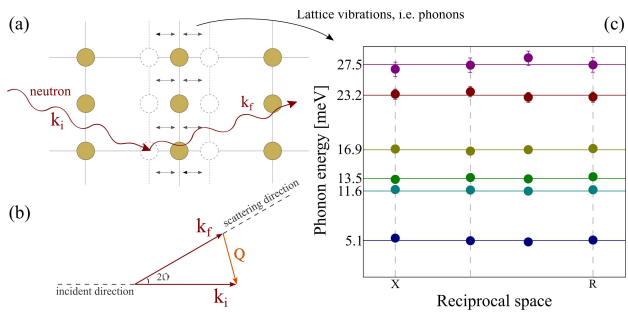}
\caption{
\textbf{Inelastic neutron scattering (INS) and dispersionless optical phonons for MAPB:} \textbf{(a)} schematic representation of lattice vibrations (phonons) interacting with neutrons with incident and final momentum, $\bf k_i$ and  $\bf k_f$ respectively.  Panel \textbf{(b)} shows kinematic condition for the conservation of momentum during the INS  experiment measuring a phonon at a momentum ${\bf Q}$ in the reciprocal (or momentum) space, from which one can deduce the phonon wavevector ${\bf q}$ from the relation  ${\bf Q}={\bf \tau} + {\bf q}$ where ${\bf \tau}$ is the nearest Bragg peak position. \textbf{(c)} Dispersionless optical phonon branches in reciprocal space as observed in MAPB with INS measurements. Two different directions of  the phonon wavevector are represented  within the Brillouin zone. 
}
\label{fig:TOC}
\end{figure}
\begin{figure*}[t]
\centering
\includegraphics[width=0.95\textwidth]{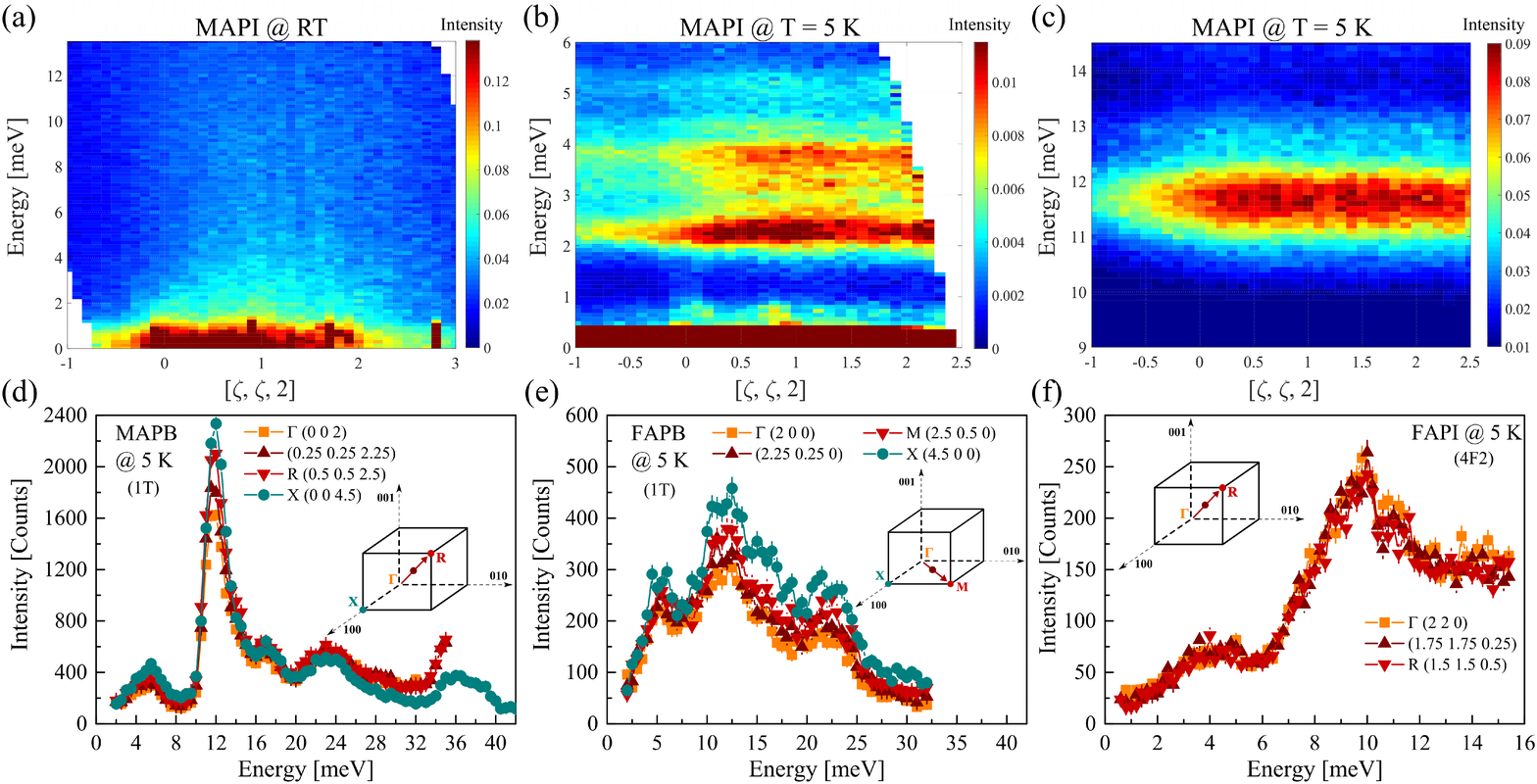}
\caption{
\textbf{Optical phonon dispersion}. Data obtained with both neutron time of flight (TOF) and triple axis spectrometer (TAS) measurements. The TOF instruments allow a mapping of optical modes in the momentum space whereas TAS focuses on particular Q-points to follow phonon modes versus momentum or energy. \textbf{(a-c)} TOF contour plots of MAPbI$_3$: \textbf{(a)} at room temperature (RT), between 0-13~meV, and at 5~K at \textbf{(b)} low energy range (below 6 meV) and \textbf{(c)} between 10-14 meV. Also at 5~K, TAS measurements along high symmetry directions, going through the reciprocal space, in \textbf{(d)} MAPB,  \textbf{(e)} FAPB and  \textbf{(f)} FAPI.  The insets of panels \textbf{(d,e,f)} show the related TAS measurement trajectories in the first Brillouin zone. The inelastic neutron scattering measurements in the orthorhombic phase (5~K) of all four perovskites, suggesting little to no dispersion of the optical phonon modes. In panel  \textbf{(d,e,f)}, error bars (of the order of the symbol size) represent one standard deviation.
} 
\label{fig:dispersion_5K}
\end{figure*} 
\indent HOPs are composed of two sub-lattices: the inorganic sub-lattice, composed of covalently bonded PbX$_3$ octahedra, and the organic sub-lattice consisting of the MA/FA molecular cations inside cuboctahedral perovskite cavities. They have been well documented to undergo a series of crystallographic transitions which differ slightly depending on the organic cation or the halide atom. It typically goes from the high-temperature cubic phase (Pm$\bar{3}$m), passing by a tetragonal phase {\color{black} (I4/mcm for  MAPbI$_3$ and MAPbBr$_3$ and  P4/mbm for FAPbBr$_3$ \cite{Schueller2017}) down to the low temperature orthorhombic phase (Pnma for all compounds except for FAPbI$_3$ where a trigonal phase P3m1 has been reported at low temperature) } \cite{Onoda-Yamamuro1990, Swainson2003, Stoumpos2013, Baikie2015, Fang2015}. In general these phase transitions originate from the tilting of the PbX$_3$ cage and the orientational ordering of MA/FA molecules, {\color{black} which is reconstructive for the low temperature transformation  in MA-based compounds  \cite{OextquotesingleLeary1970, Glazer1972,Swanson1978, Benedek2013,Letoublon2016, Beecher2016}. Note that the first order character of the low temperature phase transition is less prononced in FA-based compounds than in MA-based ones.} By inducing more complex octahedra tiltings in the perovskite lattice, organic cations may indirectly affect their electronic structure. \\ 
\indent While the molecular cations are not expected to directly contribute to the electronic band structure \cite{Borriello2008, Even2013}, they are thought to indirectly influence the electronic band edge states through the induction of distortions in the PbX$_3$ framework \cite{Even2013, Filip2014}. Effective volume in the A-site - which is increasing from K$^+$, Rb$^+$, Cs$^+$ inorganic compounds to MA$^+$ and FA$^+$ \cite{Saliba2016} - is for one affecting the Pb-X elongation. Also, the MA/FA molecules have been suggested to interact with the Pb-X network via hydrogen bonding between the ammonium hydrogens and the halide atoms, perturbing in this way the conduction band minimum, increasing diffusion length and suppressing electron-hole recombination \cite{Yin2014, Quarti2014, Li2015, Park2017}. Therefore, the interplay between the molecular groups and the inorganic network presents itself as an additional factor affecting both in the charge-recombination dynamics and the above mentioned electron-phonon interactions. \\ 

\indent On the other hand, the intrinsic anharmonicity of the perovskite lattice is expected to play a more important role \cite{Katan2018}. Lattice dynamic calculations from density functional theory ({\rm ab initio} atomic-level description) of HOPs is complicated and generally not reliable due to the hybrid nature of these compounds, where molecules perform stochastic motions at ambient temperature \cite{Katan2018}. More generally though, the strongly anharmonic character of the lattice dynamics is not taken into account in recent phonon calculations for both inorganic and hybrid compounds. Typically, the lowest energy phonons, related to Pb-halogen vibrations, are found unstable with imaginary mode frequency in almost all available theoretical calculations within the harmonic approximations \cite{Comin2016,Yang2017}. It is therefore interesting to rely on frozen phonon calculations \cite{Marronnier2018} or molecular dynamics \cite{Carignano2017} to get a first theoretical insight into anharmonic effects. \\
\indent In this work, we employed both time-of-flight (TOF) and triple-axis (TAS) inelastic neutron scattering (INS) techniques (see Methods) to investigate four different hybrid perovskite single crystals: MAPbBr$_3$, FAPbBr$_3$, MAPbI$_3$, and $\alpha$-FAPbI$_3$; from now on referred to as MAPB, FAPB, MAPI and FAPI, respectively. Complementary Raman scattering spectroscopy is also used. At low temperatures, well-defined optical phonons are observed. Mode attribution to the respective structural vibrations has been discussed and a comparison has been made between the four compounds.
We also find the optical excitations to be very weakly-propagating,  particularly in MA-based compounds. 
Moreover, the temperature dependence of the TAS spectra reveals a significant anharmonic behaviour, resulting in optical phonon overdamping at temperatures as low as 80~K. We argue for the importance of the acoustic and optical phonon coupling on the harmonicity of the lattice. \\ \\
\noindent \textbf{Results}\\

\indent  From an experimental perspective, INS allows for direct measurement of the phonon spectrum over the reciprocal space that covers the full  Brillouin zone (\textbf{Figure~\ref{fig:TOC}}), thus offering the most complete approach. Raman spectroscopy is also a very powerful and precise technique to measure optical phonons and quasielastic contributions, but restricted to the center of the Brillouin zone ($\Gamma$-point) and limited by specific selection rules. \\
\textbf{Room temperature phonon spectra}

 In \textbf{Figure~\ref{fig:dispersion_5K}}, both room temperature (RT) and low temperature (5~K) TOF and TAS measurements are presented. As a first remark,  it is necessary to cool the samples down to the lowest temperature to observe well-defined optical phonon branches. Indeed, the INS spectrum recorded at room temperature in MAPI shows no well-defined optical phonon modes at any energy range as a result of all modes being overdamped in all momentum points (\textbf{Figure~\ref{fig:dispersion_5K}.a}).  Only low energy acoustic phonons can be actually identified (see also Supplementary Figure~1), in agreement with the report of RT acoustic phonons in the same four HOP compounds \cite{Letoublon2016,Ferreira2018}.  Above  $\sim$3 meV, the acoustic branches vanish \cite{Ferreira2018} as observed in the broadening of the acoustic branches at the zone boundary in deuterated MAPI \cite{Toney} and in MAPbCl$_3$ \cite{Songvilay2018}.  The absence of optical phonons at RT is also observed in TAS spectra of the other three systems in the Supplementary Figure~2, all of which is in line with a previous report on MAPI \cite{Li2017}.  {\color{black} Accordingly, only a broad inelastic contribution was  observed around 12 meV  in MAPB at RT and a tentative attribution of the various phonon modes inside this bundle was only possible through the numerical fitting of the experimental spectrum by a superposition of  damped harmonic oscillators \cite{Letoublon2016}.} This is also consistent with the high temperature Raman spectrum of MAPbBr$_3$ 
(vide infra) \cite{Yaffe2017}. The dispersive optical phonon modes reported at RT by inelastic X-Ray scattering in MAPI and MAPB \cite{Comin2016}, with its broader energy resolution (\rm{i.e.} $\sim$ 1.5 meV)  {\color{black} and its Lorentzian instrumental shape,} are then put into question.

\textbf{Lack of dispersion in momentum space}

\begin{figure}[t!]
\centering
\includegraphics[width=1\linewidth]{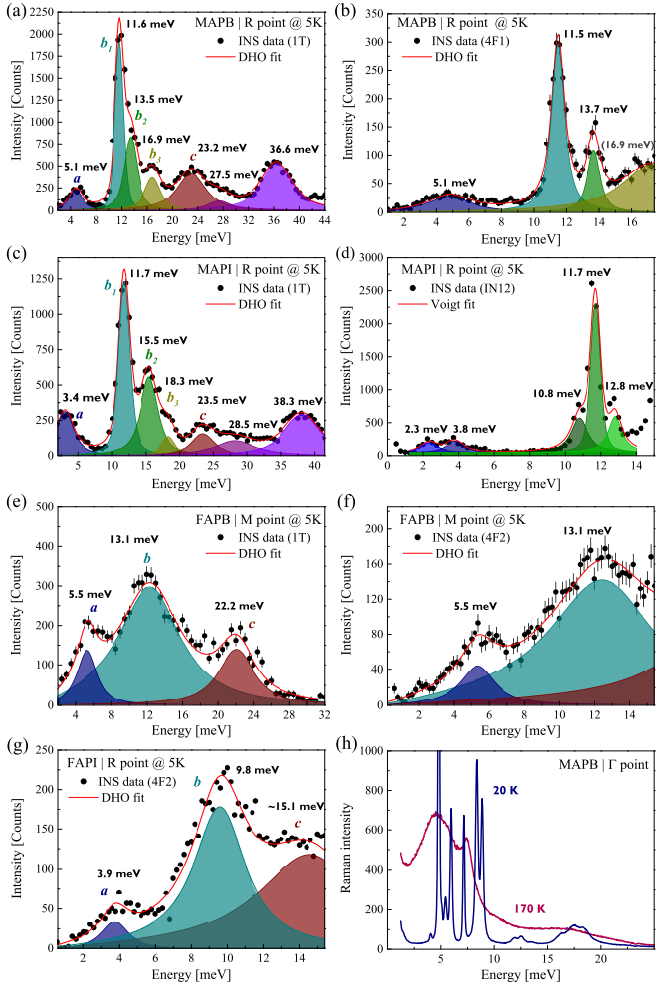}
\caption{
\textbf{Low temperature optical phonon spectra.} 
Triple axis spectrometer (TAS)  neutron spectra measured at 5~K, at the R bragg reflection (1/2, 1/2, 3/2) of \textbf{(a,b)} MAPB, \textbf{(c,d)} MAPI,  \textbf{(e,f)} at the M point (5/2, 1/2, 0) of FAPB and \textbf{(g)} at the R point of FAPI.  For each compound, each row shows the measurements either using a thermal \textbf{(a,c,e)} or a cold \textbf{(b,d,f)} neutron instrument, with the exception of FAPI only using cold TAS  \textbf{(g)}.
 The energy resolution using TAS varies considerably with experimental conditions and the neutron energies. 
It is broader on thermal instrument (1-3~meV from 0 to 40~meV energy transfer) than on cold neutron instruments (0.2-0.4~meV from 0 to 15~meV energy transfer). The experimental TAS data (black scatter points) is fitted (full red line) with a sum of damped harmonic oscillators (see methods) and is presented with a removed constant background. Individual fitted peaks are labelled (filled-coloured area). \textbf{(h)} Comparison of the Raman scattering responses above (170~K) and below (20~K) the orthorhombic-tetragonal transition. Error bars (sometimes smaller than the symbol size) represent one standard deviation.
}
\label{fig:optical-modes-5K}
\end{figure}

 In order to be able to observe sharp optical phonon features, recording neutron spectra at 5~K (\textbf{Figure~\ref{fig:dispersion_5K}.b-f)} is necessary as cooling down to the lowest temperature substantially reduces phonon damping. In MA-based HOP compounds, the phonons are typically resolution-limited in energy at 5~K, whereas in FA-based systems the phonon features remain broad even at 5~K (see \rm{e.g.} \textbf{Figure~\ref{fig:optical-modes-5K})}. {\color{black} Showing relatively broad peaks that apparently represent scattering from a bundle of modes, it should be emphasized that INS technique is incapable of detecting a  specific phonon dispersion in such closely packed optical modes which might overlap. However, this experimental limit depends greatly on the instrumental energy resolution which is very much improved for cold neutron spectrometers (see Fig. \ref{fig:optical-modes-5K} caption). }

Having said that, one observes that all phonon modes (or mode bundles) show little to no dispersion as shown by TOF in MAPI (\textbf{Figure~\ref{fig:dispersion_5K}.b,c}). The low temperature TAS measurements along high symmetry directions ($\bf \Gamma \rightarrow M$ or $\bf \Gamma \rightarrow R$) in MAPB, FAPB and FAPI show no dispersion either within a 0.1~meV error, and only the amplitudes of the modes vary (\textbf{Figure~\ref{fig:dispersion_5K}.d,e,f}).  That shows that optical excitations are almost non-propagating as summarized in  \textbf{Figure~\ref{fig:TOC}.b} for MAPB, 
suggesting a localized character in real space. 
\begin{figure}[t]
\centering
\includegraphics[width=0.9\linewidth]{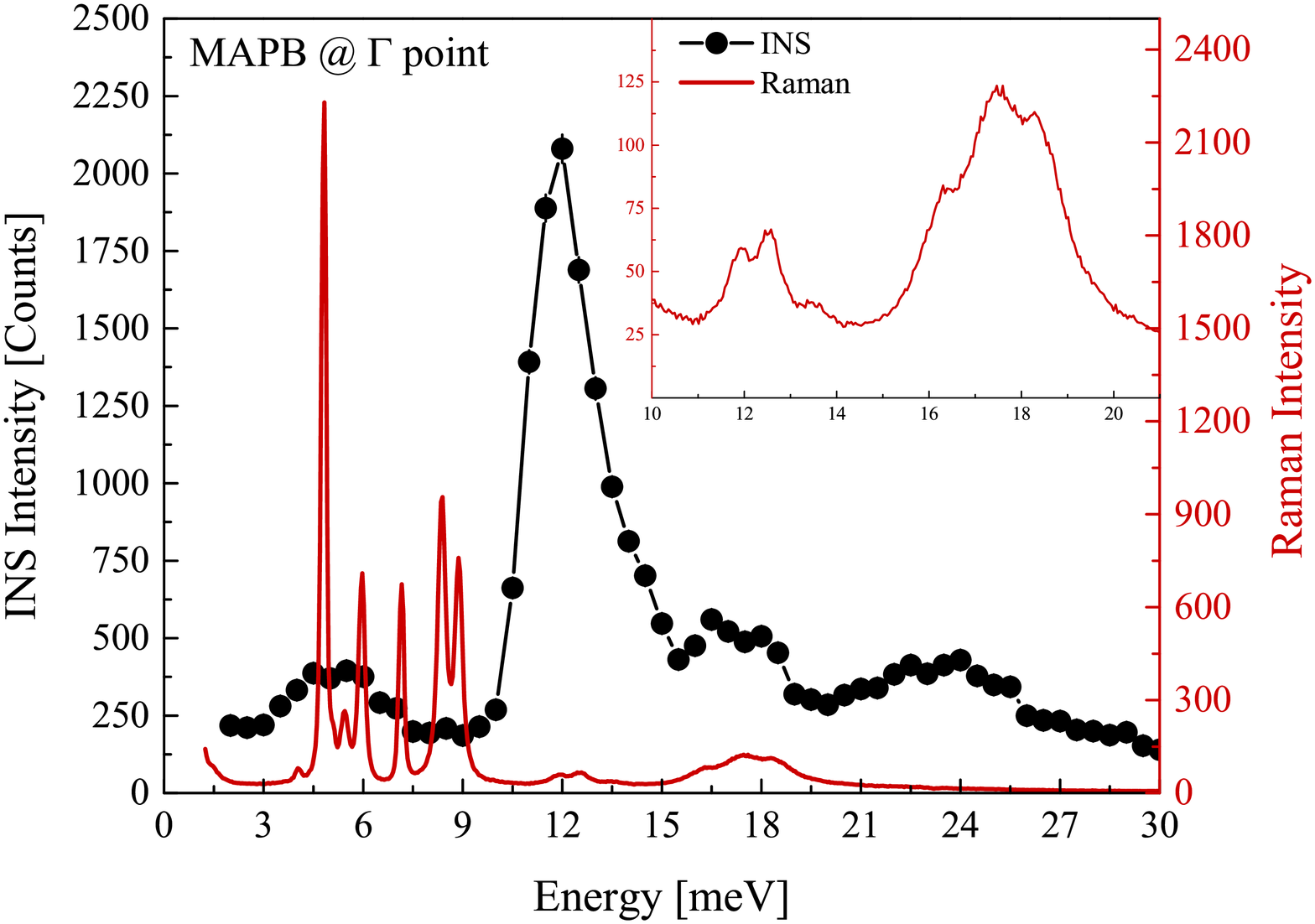}
\caption{
\textbf{Raman response of MAPB at low temperature.} Comparison of the triple axis spectrometer (TAS) neutron data using the thermal beam instrument 1T  (5~K) and Raman scattering (20~K) optical phonon spectra in MAPB, at the $\Gamma$ point ($\equiv (0,0,2)$ Bragg position for INS). Inset shows a magnification of the Raman scattering spectra in the 10-20~meV range.
The energy resolution of Raman scattering is here 0.125~meV compared to the energy resolution of $\sim$1-2~meV for a thermal TAS. Error bars (smaller than the symbol size) represent one standard deviation.
}
\label{fig:Raman_vs_INS}
\end{figure}

However,  we need to temper this observation. Indeed, as pointed out in Supplementary Note 1, the neutron intensities correspond to the sum of coherent and inchorent cross-sections. In case of incoherent scattering, the phonon momentum information is lost. As hydrogen atom has a large incoherent scattering length, the phonon peaks involving hydrogen would roughly correspond to the incoherent cross-section. In contrast, the phonon peaks from the PbX$_3$ cage are  related to the coherent cross-section, that typically corresponds to the lowest energy bundle  \textit{a} (as discussed below) showing as well no dispersion (see {\rm e.g.}  the TOF data in MAPI, \textbf{Figure~\ref{fig:optical-modes-5K}.b}). Furthermore, even for incoherent scattering, some information can be obtained from the sharpness of the measured peaks. Energy resolution-limited peaks, as we observed for MA based HOP,  can only be accounted for by dispersionless phonons even for incoherent scattering. It is interesting to compare our data with the neutron scattering studies of HOP powder samples where the ${\bf q}$ dependence is also lost by averaging over all orientational directions. 
It appears that the powder sample spectra of MAPB \cite{Swainson2015} and MAPI \cite{Schuck} also show sharp phonon modes at the same energy than in our experiments,  {\color{black}  suggesting once more the dispersionless nature of these phonon modes. The neutron powder sample spectrum observed in  MAPbCl$_3$\cite{Schuck} does not show any sharp features in clear contrast with MAPI, suggesting again that optical phonons exhibit no dispersion in MA-based HOPs. }

 {\color{black}  It should be further stressed that the lowest energy mode at 5 meV in MAPB \cite{Swainson2015} 
is found at the same position in protonated and deuterated samples,  showing that it is not related to a molecular vibration. The lowest energy mode is therefore necessarilly associated to the coherent neutron cross section, corroborating the lack of dispersion in MA-based HOPs of the lowest frequency modes. In contrast, } the observation of broader phonons modes in FA-based HOPs (see below) may question the dispersionless nature in FAPB and FAPI (for instance, the modes  above $\sim$7 meV in \textbf{Figure~\ref{fig:optical-modes-5K}.e-g} involving molecular vibrations). Indeed, the origin of broader phonon peaks can be either due to a larger damping or related to moderately dispersive phonons in FA-based HOP.

 The observation of dispersionless phonon is opposed to what has been previously predicted using density functional perturbation theory in the frozen-phonon approximation in MA-based HOPs \cite{Comin2016,Yang2017}, where dispersive phonon branches were computed. It is worth to emphasize that this lack of dispersion concerns both longitudinal and transverse optical phonons.  {\color{black} For FA-based HOP,  due to the large broadening still observed at low temperature, slight phonon dispersion associated to  crossing of  branches inside the three bundles can not be completely ruled out.} \\
\begin{table*}[t!]
\def\arraystretch{1.7}
\centering
\tiny
\caption{Energies of the optical phonon modes measured at 5~K by inelastic neutron scattering (INS) and 20~K by Raman scattering. Phonon lines (or bundles of modes for INS) are fitted by a sum of damped harmonic oscillators. Energies are given in meV. The error bars of the phonon energies obtained from the fit of INS spectra are 0.1-0.2 meV unless stated otherwise. The error bars of the phonon energies obtained from Raman scattering are 0.05 meV for low energy, 0.2 meV for the medium energy range and 0.4 meV for the high energy range.  {\color{black} Mode characters are indicated as they are discussed in the main text.}}
\label{tab:list_modes}
\begin{ruledtabular}
\begin{tabular}{ccccccc}
\multicolumn{1}{c}{\multirow{2}{*}{Energy}}  &  \multicolumn{1}{c}{\multirow{2}{*}{Mode character}}     & \multicolumn{2}{c}{MAPB}                                                                                                                          & \multicolumn{1}{c}{MAPI}                                                & \multicolumn{1}{c}{FAPB} & \multicolumn{1}{c}{FAPI}                              \\ \cline{3-7} 
\multicolumn{1}{c}{}                                  &              & INS                                                        & \multicolumn{1}{c}{Raman}                                                            & \multicolumn{1}{c}{INS}                                                 & \multicolumn{1}{c}{INS}  & \multicolumn{1}{c}{INS}                               \\ \hline
\begin{tabular}[c]{@{}c@{}}Low \\ ($<$~10~meV)\end{tabular} & 
\begin{tabular}[c]{@{}c@{}} PbX$_3$ rocking\\ and bending  \end{tabular}  & 5.1                                                     & \begin{tabular}[c]{@{}c@{}}4.1 \ $|$ \ 4.8 \ $|$ \ 5.1 \ $|$ \ 5.5 \\  5.9 \ $|$ \ 7.2 \ $|$ \ 8.4 \ $|$ \ 8.9\end{tabular} & \begin{tabular}[c]{@{}c@{}}2.3  \\ 3.8  \end{tabular}                        & 5.5              & 3.9                     \\ \hline
\begin{tabular}[c]{@{}c@{}}Medium \\ (10-20~meV)\end{tabular}  &  
\begin{tabular}[c]{@{}c@{}} PbX$_3$ stretching \\ molecular rattling \end{tabular}  & \begin{tabular}[c]{@{}c@{}}11.6 \ $|$ \ 13.5 \\ 16.9\end{tabular} & \begin{tabular}[c]{@{}c@{}}11.9 \ $|$ \ 12.6 \ $|$ \ 13.5 \\ 16.3 \ $|$ \ 17.3 \ $|$ \ 18.3\end{tabular}       & \begin{tabular}[c]{@{}c@{}}10.8 $\pm$0.3$\ |$ \ 11.7 \ $|$ \ 12.8 \\ 15.5 \ $|$ \ 18.3\end{tabular} & 13.1                      & \begin{tabular}[c]{@{}c@{}}9.8$\pm$0.3\\  15.1$\pm$3\end{tabular} \\ \hline
\begin{tabular}[c]{@{}c@{}}High \\ ($>$~20~meV)\end{tabular} &
\begin{tabular}[c]{@{}c@{}} Molecular\\ librations  \end{tabular}  & \begin{tabular}[c]{@{}c@{}}23.2 \ $|$ \ 27.5$\pm$0.4\\ 36.6$\pm$0.4\end{tabular} & \begin{tabular}[c]{@{}c@{}}21.8\\ 40.8\end{tabular}                                   & \begin{tabular}[c]{@{}c@{}}23.5 \ $|$ \ 28.5$\pm$0.6\\ 38.3\end{tabular}               & 22.2                      &                                                        \\ 
\end{tabular}
\end{ruledtabular}
\end{table*} 
\indent Raman scattering spectroscopy has been performed as well  in the MAPB single crystal as a control experiment (see Methods).  \textbf{Figure~\ref{fig:Raman_vs_INS}} shows the comparison of the low temperature (20~K) spectra of MAPB, between INS on a thermal instrument and Raman scattering (that agrees with a previous Raman scattering report \cite{Guo2019}). Raman spectra are recorded at very low \textit{q}, whereas the INS experiments are measured at a Bragg peak position ${\bf Q}=(0,0,2)$, both are therefore probing optical phonons at the $\Gamma$ point in the Brillouin zone. 
The energy positions of the optical phonon modes here identified from INS and Raman are consistent, especially the Raman mode bundles around 5~meV, 12~meV and 18~meV (inset).
{\color{black} However, the modes below 9~meV are significantly broader and  less intense in INS than in Raman scattering.
This  may be attributed to a poorer energy resolution in INS for thermal TAS (1-2~meV) as compared to Raman (0.125 meV) as well as to the different scattering efficiency between both techniques.}
 \textbf{Figure~\ref{fig:dispersion_5K}} shows as well that each optical phonon mode can be defined by almost the same energy for any Q-point of the Brillouin zone, although mode broadening related to instrumental resolution is more important for neutron scattering than for Raman scattering. \\
\textbf{Optical phonons in the orthorhombic phase}

 Systematic low temperature INS experiments were performed with medium and high energy resolutions from, respectively, thermal and cold beams (see Methods). The optical phonon spectra (obtained at either the M point or R point) for all four compounds are presented in \textbf{Figure~\ref{fig:optical-modes-5K}}. Each compound studied undergoes structural distortions at low temperature, giving rise to atomic superstructures at either or both the M and R points. At low temperature, phonon spectra have been recorded at both momentum points. As no noticeable difference can be observed between these spectra (related to the lack of dispersion of the phonon modes), we here report the phonon spectrum measured at the R point for most compounds, except for FAPB, where it is shown at the M point (see also Supplementary Figure~3). As already stated, a series of prominent optical phonon bundles of modes are observed at different energy ranges. Each phonon mode is usually accounted for by a damped harmonic oscillator (DHO). Therefore, a model with a sum of DHOs convoluted by the spectrometer energy resolution is used to describe the neutron spectra, including a constant background (see Methods).  The neutron spectra of \textbf{Figure~\ref{fig:optical-modes-5K}} have been fitted using \textbf{Eq.~\ref{DHO}} and the obtained phonon energies are summarized in \textbf{Table~\ref{tab:list_modes}}.\\ 
\indent For a first analysis, we define three energy ranges where the different optical modes are located. These modes have been arbitrarily labelled  as low (2-10~meV), medium (10-20~meV) and high ($>$ 20~meV). More specifically, in MAPB we find a low-energy mode at 5~meV; an intense peak at 11.6~meV followed by two smaller shoulders at 13.5 and 16.9~meV; and relatively broader bands at 23.2, 27.2 and 36.6~meV (\textbf{Figure~\ref{fig:optical-modes-5K}.a,b}). These results match well with a previous INS experiment on a MAPB powder sample \cite{Swainson2015}.\\
\indent Similarly in MAPI, we obtained two low-energy modes located at 2.3 and 3.8~meV, an intense peak centered at 11.7~meV surrounded by two smaller shoulders at 10.8 and 12.8~meV followed by two bundles at 15.5 and 18.3~meV, and at higher energy range, other ones at 23.5, 28.5 and 38.3~meV (\textbf{Figure~\ref{fig:optical-modes-5K}.c,d}). These again are directly comparable with previous INS studies \cite{Druzbicki2016, Li2017} where two additional small modes in the 2-5~meV range ($\sim$3.1 and $\sim$4.3~meV) were reported.\\
\indent In FAPB, we note a broader central feature dominating the medium energy range (\textbf{Figure~\ref{fig:optical-modes-5K}.e,f}), while three distinct bundles can be identified at 5.5, 13.1 and 22.2~meV. These results on FAPB are in line with the ones observed in MAPB, although much broader phonon bundles are systematically observed in FAPB compared to the more numerous (and narrower) modes that appear for MAPB. These two compounds share the same space group (Pnma) at low temperature although the static structural distortions characteristic of the orthorhombic phase are smaller in FAPB \cite{Schueller2017}. This may be connected to a different dynamics of the FA cation and its coupling to the perovskite lattice, by comparison to the MA cation (vide infra) \cite{Carignano2017}.\\
\indent Finally, in FAPI - a sample requiring specific procedures and a proper storage to avoid the transformation to its yellow non-perovskite $\delta$-phase at high temperature before cooling \cite{Zhumekenov2016} and, therefore, scarcely studied at such a low temperature in the literature - we could only  perform measurements on the 4F2 cold neutron spectrometer (\textbf{Figure~\ref{fig:optical-modes-5K}.g}). Nevertheless, in the accessible energy range, we detect large optical phonons at 3.9, 9.8 and $\sim$15.1~meV. PL studies on single CQDs \cite{Fu2018} have identified exciton side-bands at 3.2, 7.8 and 15.4~meV, which match rather well with our TAS results on FAPI. The intermediate mode seemingly exhibits a slight discrepancy, but as reported in \cite{Fu2018}, it undergoes temporal fluctuations under high energy excitation, and is roughly spread between 7.5 and 12.5~meV. Again, our TAS results on FAPI show a significantly broad profile, as in FAPB.\\
\indent According to nuclear magnetic resonance (NMR) measurements~\cite{Kubicki2017}, the broader nature of the optical modes of FA-based compounds, compared to their MA-based counterparts, may be attributed to the fact that the FA reorientation in FA-containing materials is faster than that of MA in the MA-based perovskite, despite the fact that FA is larger than MA. 
This has an impact on the charge carriers lifetime in these compounds. In addition, the acoustic density of states is located at lower energy in FA-based compounds as compared to the MA-based  \cite{Ferreira2018}, thus leading to enhanced scattering between acoustic and optical phonons and related increased anharmonicity. \\
\indent In \textbf{Figure~\ref{fig:optical-modes-5K}.h} we further compare Raman scattering spectra in MAPB, above and below the orthorhombic to tetragonal first order transition $\sim$150~K. When passing the phase transition and on cooling down to 20~K one notices a considerable narrowing of the phonons lines below 10~meV, while higher frequency vibrations evolve much more smoothly and continuously. Furthermore, phonon modes are split at low temperature below 10~meV (\textbf{Figure~\ref{fig:Raman_vs_INS}}). These additional Raman scattering modes (listed in \textbf{Table~\ref{tab:list_modes}})  result from the band folding in the low temperature phase, induced by the structural distortion caused by the phase transition. \\
\indent Overall, a commonality between the different perovskite systems has now been identified.
 By cross-referencing our results with the above mentioned previous works and other literature on the subject, we can assign the identified features of the low temperature spectra to the respective vibrational modes. In recent lattice dynamics calculations, Ponce \rm{et~al.} \cite{Ponce2019} predict five modes for MAPI at 3, 4.3, 10.2, 14.4 and 21~meV. A-site displacement (\rm{i.e.} rattling of the organic molecule within the cage) is said to be responsible for the peak at 10.2~meV, while scattering at 21~meV results from libration motions of the organic cations. On the other hand, the modes at 3, 4.3 and 14.4~meV are claimed to be related to the inorganic sub-lattice and arise from the rocking, bending and stretching motions of the PbI$_3$ network, respectively, the latter one involving hybridization with organic cation motions. The calculated modes match well with our measurements and previous INS studies \cite{Druzbicki2016, Li2017,Schuck} and the identifications made based on associated density functional theory  (DFT) calculations \cite{Druzbicki2016}.\\
\indent Regarding MAPB, one is compelled to draw parallel conclusions with those made about MAPI. Besides, in the work by Swainson \rm{et~al.} \cite{Swainson2015} on MAPB, the 5~meV mode is again associated to vibrations of the PbX$_3$ network. Furthermore, their comparative study between non-deuterated and deuterated samples clearly evidenced the influence of the organic cation motions on the modes at 11.5 and 13.7~meV.\\
\indent In the PL study on FAPI CQDs \cite{Fu2018}, the authors compared their study with the theoretical predictions and near-infrared spectroscopic measurements on MAPI \cite{Perez-Osorio2015} and, as a result, ascribe the observed side-bands to LO phonon modes related to bending (3.5~meV) and stretching (15~meV) motion of the PbI$_3$ cage, and to rigid-body motions of FA cations (11~meV). 
%

%
\begin{figure*}[t]
\centering
\includegraphics[width=0.7\textwidth]{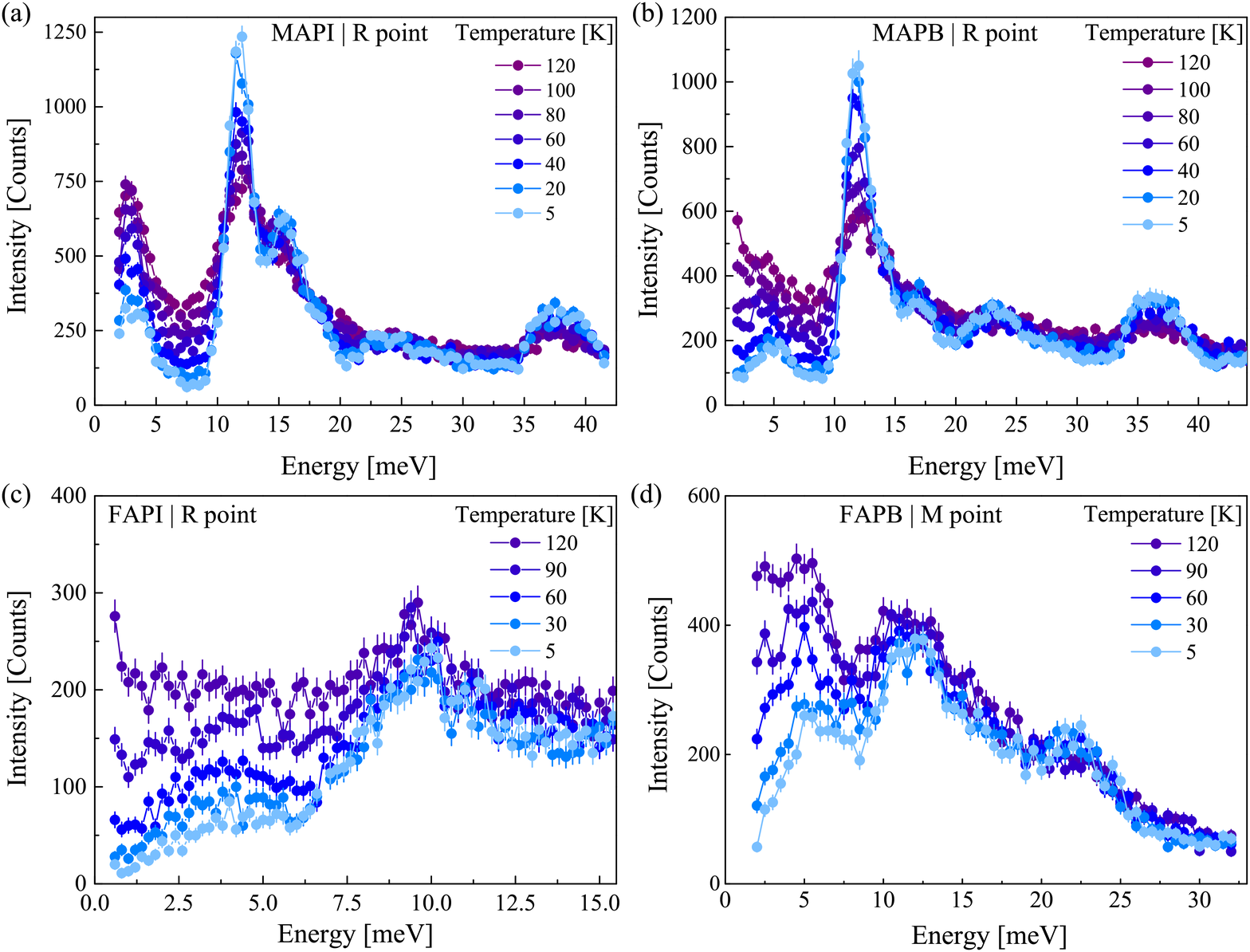}
\caption{
\textbf{Temperature dependence of neutron spectra}. Optical phonon spectra as a function of temperature, at the R point of \textbf{(a)} MAPI, \textbf{(b)} MAPB, \textbf{(c)} FAPI, and \textbf{(d)} at the M point of FAPB. Measurements in \textbf{(c)} were performed on a cold neutron source, hence the difference in energy range, as compared to the \textbf{(a), (b)} and \textbf{(d)} scans (thermal instrument).  Error bars (sometimes smaller than the symbol size) represent one standard deviation.
} 
\label{fig:INS-Temp}
\end{figure*} 
\begin{figure}[t]
\centering
\includegraphics[width=0.9\linewidth]{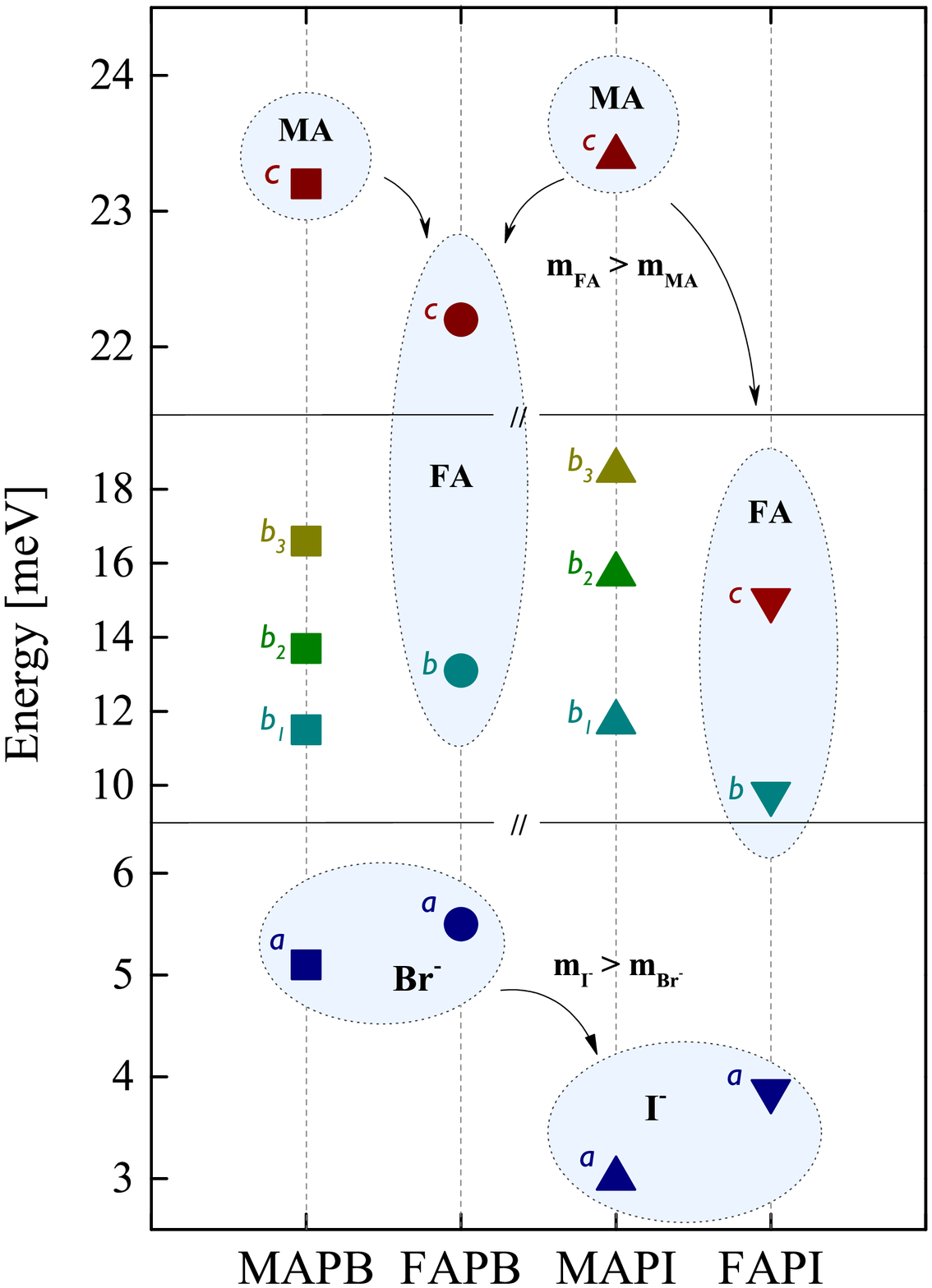}
\caption{
\textbf{Main phonon bundles comparison.} We illustrate the relative energy positions of the optical phonon bundles \textit{a}, \textit{b} and \textit{c}, between MAPB, FAPB, MAPI and FAPI. The figure highlights an obvious influence of the molecular and/or the halide component, on the respective vibrational modes and the associated optical phonon energy.
}
\label{fig:OptModes-vs-Perov}
\end{figure}

\indent As far as the mode at $\sim$37~meV is concerned, there is a consensus that it originates from organic molecular vibrations \cite{Druzbicki2016, Park2015, Quarti2013}, although there is a debate about the exact nature of the involved atomic motions. Park \rm{et~al.} \cite{Park2015} describe MA vibrations involving MA wagging, MA rotation and MA-MA stretch. Quarti \rm{et~al.} \cite{Quarti2013}, as well as Dr\.uzbicki \rm{et~al.} \cite{Druzbicki2016}, suggest a torsional MA vibration (also called disrotatory vibrations) that involves the terminal NH$_3$ and CH$_3$ moieties, which is also in line with the vibrational mode found near 300~cm$^{-1}$ (\rm{i.e.} 37.2~meV)  in isolated MA calculations \cite{Quarti2013, Zhou2018}.\\

\noindent \textbf{Anharmonic behaviour of lattice dynamics} 

As mentioned above, anharmonicity of the hybrid halide perovskite lattice is expected to play a considerable role in electron-phonon interactions. In \textbf{Figure~\ref{fig:INS-Temp}} we show the temperature behaviour of the INS optical phonon spectra of the four compounds, up to 120~K. The anharmonicity manifests already at low temperatures well below RT, where optoelectronic devices and solar cells are usually operating. The increased phonon damping is observed in \textbf{Figure~\ref{fig:INS-Temp}} above $\sim$30~K, together with an increase of the low energy quasi-elastic signal and a reduced phonon intensity. The weaker phonon intensity upon warming corresponds to a decrease of the phonon Debye–Waller factor, meaning  the increase of atomic mean displacements. However, consistent DHO fitting of the temperature data has also proven difficult due to the strong anharmonic behaviour. These effects lead to a rather quick overdamping of certain phonon modes at temperatures as low as 80~K, making it difficult to properly assign the exact contribution of each phonon mode to the overall spectra at higher temperatures.\\ \\
\noindent \textbf{Discussion}\\
The low temperature INS (and Raman) optical phonon spectra in our four HOP single crystals, reveal a number of characteristic features down to very low energy. It is now accepted that carriers  scattering in HOPs is dominated by Fröhlich coupling between charge carriers and optical phonon modes. This phonon scattering is believed to be the key fundamental factor in establishing the intrinsic limit of the charge carriers mobility. More specifically, it has been recently suggested that this limit is set by the lower energy LO modes (3-20~meV) \cite{Ponce2019}. Therefore, the observed presence of such modes, common in all four compounds, seems to be the reason for the relatively low mobilities compared to classical inorganic semiconductors like Si and GaAs. Nevertheless, our experimental study clearly indicates that a missing ingredient of current attempt to reproduce the observed temperature dependence of charge carriers mobilities in HOPs is related to the underlying harmonic or quasi-harmonic assumptions of phonon modelling. Defining optical phonons as well-defined quasi-particles above 80-100~K is actually questionnable according to the present experimental observations. Therefore the apparent discrepancy between the experimental acoustic-like temperature dependence of carriers mobilities and the expected dominant process (Fröhlich interaction) may be an unexpected consequence of the HOPs lattice softness.
 \\
\indent Three different energy ranges have been identified  in the spectra and gathered the modes or bundles of modes for each compound in three categories (\textit{a}, \textit{b} and \textit{c}), as can be seen in Figure~\ref{fig:optical-modes-5K}. The following conclusions are drawn from our experimental data and based on earlier literature. Modes in the low-energy range (below 10~meV) are associated with vibrations of the PbX$_3$ network, mainly rocking and bending. As for the intermediate energy range, between 10 and 20~meV, a series of mutually coupled modes are observed which arise from both the organic and inorganic sub-lattices and, therefore, show a highly hybridized nature. Here, a prominent feature is the stretching of PbX$_3$, which is predicted to be coupled with the organic sub-lattice. An additional proof of this coupling comes from the comparison between non-deuterated and deuterated samples \cite{Swainson2015}. This further corroborates the increased broadening observed in FA-based systems, especially in this medium energy range. This enhanced broadening is consistent with the difference in the dynamics of the FA and MA cations previously  reported by NMR~\cite{Kubicki2017}, and molecular dynamics simulations \cite{Carignano2017}. It has also been shown by PL studies that the phase transition down to the lowest temperature is much smoother (weaker distortion) in FA-based compounds \cite{Schueller2017, Fang2015} than in MA-based ones \cite{Fang2015a}. However, it should be stressed that the broadening in the FA compounds can be affected  by the possible dispersion of bundles \textit{b} and \textit{c} due to the incoherent nature of these peaks in INS experiments. 
Also, still  for the bundle  \textit{b}, a phonon related to the A-site displacement is identified at least in MAPI. 
Above 20~meV, optical features are essentially a result of molecular motions, although there could be some inorganic contributions to the cation librations of bundle \textit{c}, at least for MA-containing perovskites. Meanwhile, direct comparison between the MAPB and FAPB spectra leads us to believe that the same mechanisms are possibly at play in the latter. It is apparent, that in FA-based compounds there is significantly more contribution stemming from the coupling between inorganic and organic sub-lattices, as a result of the more hybridized nature of the lowest energy modes.\\
\indent In \textbf{Figure~\ref{fig:OptModes-vs-Perov}}, the relative energy shifts of the labelled phonon bundles, between the four compounds, is presented. One can see that the low-energy bundles \textit{a} have lower energy in compounds that contain I$^-$ as opposed to Br$^-$. Likewise, there is a significant decrease in energy in the medium/high energy bundles when coming from MA-based compounds to FA ones. Reminding that the phonon energy is generally proportional  to the square root of the inverse of atomic mass (E $\sim  1/\sqrt{M}$) and that FA$^+$ and I$^-$ are heavier than MA$^+$ and Br$^-$ respectively, we can conclude that such trends agree with the attribution regarding the origin of the respective modes (see Supplementary Note 2). Actually, the relative shifts observed here, \rm{i.e.} lower energy in FA$^+$ and I$^-$ based compounds when comparing with MA$^+$ and Br$^-$, are also consistent with the smaller elastic constants obtained in our previous INS study \cite{Ferreira2018}. \\
\indent In 3D hydrid peroskites, substantial interactions between the organic and Pb-halogen neighbouring networks are typically expected, however, besides being located in low energy range, the optical lattice excitations measured here appear to be basically non-propagating (\textbf{Figure~\ref{fig:dispersion_5K}}), exhibiting nearly no dispersion in the Brillouin zone, and thus contradicting existing reports on phonon simulations based on the harmonic approximation. Similarly to phonon modes in thermoelectric chlarates \cite{clathrate}, strong anharmonic phonon–phonon scattering processes may lead to a series of anticrossings flattening phonons dispersions. For instance, this strong hybridization between phonons would typically involve rattling of the organic molecule within the PbX cage. Furthermore,  rattling phonon modes are generally associated with anharmonicity and lower thermal conductivities \cite{Rowe1995, Voneshen2013} as it has been discussed through a crystal-liquid duality of HOPs \cite{ Miyata2017a}.

%
\indent While the anharmonicity of halide perovskites leads to low frequency acoustic phonons, characteristic of a soft lattice \cite{Ferreira2018}, it occurs for optical phonons through an overdamped behaviour over the entire Brillouin zone at room temperature, and weakly dispersing branches at low temperature. The overriding anharmonic character of optical phonons specific to halide perovskites is most probably an important missing link for a proper account of the leading Fröhlich carrier-phonon interaction for this class of soft semiconductors. The size of the cation and the nature of the halogen are additional features known to strongly influence the softness of the lattice \cite{Ferreira2018}. They also have a direct impact on the damping of the optical modes. Therefore, this suggests that the coupling between acoustic and optical phonons may play a role in the harmonicity of the lattice, besides non-linear coupling between optical phonons. \\
\indent The lack of dispersion of such low optical modes overlapping with the upper part of the acoustic phonon dispersions is expected to be at the origin of several specific physical properties of HOPs.
 We first note a significant anharmonic behaviour that manifests itself in phonon overdamping at temperatures well below the ones used for operating optoelectronic devices and solar cells. Therefore the current modelling of charge carriers mobilities based on a quasi-particle picture for low-energy optical lattice modes is questionable. Together with an apparent correlation between lattices softness (\rm{i.e.} elastic constants) and optical phonon energy/frequency, our results point to the influence of acoustic and optical phonon coupling on the harmonicity of the lattice and the Fröhlich interaction between charge carriers and optical phonons. Moreover, this coupling occurs as well in the anharmonicity of the upper part of the acoustic phonon branches.\\
%
\indent In conclusion, we present here a extensive comparison of optical phonon excitations in four different hybrid organolead perovskite compounds. INS and complementary Raman scattering measurements revealed various features of the  low temperature phonon spectra, that we have assigned to possible structural vibrations.  The dispersionless nature of these optical modes is a first characteristic feature of HOPs. 
The optical phonons overdamping upon warming for temperatures as low as $\sim$ 80~K  demonstrate the strong anharmonicity in these materials. We believe that the following experimental report could serve as a solid base in future theoretical calculations and modelling, for improved mode assignment and understanding of the electron-phonon interactions, especially for FA-based compounds where measurements of their optical phonon spectrum have been lacking.\\ \\
\noindent \textbf{Methods}\\
\footnotesize{
\noindent \textbf{Sample preparation} 

Single crystals of four different hybrid lead halide perovskites have been grown by the inverse temperature crystallization (ITC) method (see Supplementary Note 3). The perovskite compounds are: MAPbBr$_3$ (MAPB), FAPbBr$_3$ (FAPB), MAPbI$_3$ (MAPI), and $\alpha$-FAPbI$_3$ (FAPI). MA and FA stand for methylammonium  and  formamidinium molecules, respectively. All single crystals {\color{black}  of typical size of 200 mm$^3$ for all compounds (except for FAPI where only a volume of $\sim$ 50 mm$^3$  could be achieved)}, were  synthesized at the Institut des Sciences Chimiques de Rennes (ISCR). \\ \\
\noindent \textbf{Inelastic neutron scattering spectroscopy} 

Inelastic neutron scattering (INS) measurements were conducted using both Triple-Axis Spectrometers (TAS) and Time-of-Flight Spectrometers (TOF). On all TAS instruments, monochromators and analysers were made from the 002 reflection of Pyrolithic graphite (PG). Cold (below $\sim$15~meV) and thermal TAS have been used to cover the full energy range of the phonon spectrum in the three hybrid lead halide perovskites, MAPI, MAPB and FAPB. On 4F1/4F2 and IN12 TAS located (on cold-neutron sources), respectively, at the reactor Orph\'ee/Laboratoire L\'eon Brillouin (LLB), in CEA Saclay and at the Institut Laue-Langevin (ILL) in Grenoble, a constant final neutron wave vector of $\rm k_f = 1.55 \AA^{-1}$ was utilized with a beryllium (Be) filter to remove high-order neutrons in the beam. In addition for the measurement on IN12, a velocity selector was used to remove neutrons with high order harmonics from the incident beam. At the thermal TAS 1T at LLB, a constant final neutron wave vector of $\rm k_f = 2.662 \AA^{-1}$ was used with a pyrolytic graphite filter to remove neutrons with high order harmonics. The energy resolution of cold TAS goes from $\sim$0.2 to 0.4\,meV for an energy transfer ranging from 0 to 15~meV, while for the thermal TAS (1T) it goes from $\sim$1 to 3~meV for an energy transfer ranging from 0 to 40~meV. \\
\indent In MAPI, mapping of the phonon spectrum has been performed using the TOF instrument IN5 at the Institut Laue-Langevin (ILL) in Grenoble with an incident neutron wavelength of $\lambda= 2$\AA\ ($\equiv$ 3.14~\AA$^{-1}$), corresponding to an energy resolution which varies from 1.15 meV at elastic position to 0.8 meV at 15 meV energy transfer. The four-dimension $S({\bf q},\omega)$ data measured on IN5 were reduced and visualized using the Horace software suite. Cuts shown in \textbf{Figure~\ref{fig:dispersion_5K}.a-c} were made along [hh2] direction, with an integration over -0.1 $< l <$ 0.1 (i.e. [002+l]) and -0.1 $< \eta < $ 0.1 in [$\eta$,-$\eta$, 0].  \\
\indent All samples have been attached to the cold head of the cryogenerator at LLB or a cryostat at ILL, reaching a low temperature of 5~K where all these hybrid organolead perovskites are in the orthorhombic phase. However, throughout the present manuscript, the Miller indices refer to the high temperature cubic phase of the perovskite lattice. Samples were mounted in a scattering plane such that the high symmetry reciprocal directions [001] and [110] were within the horizontal plane, except for FAPB which mounted with directions [100] and [010] within the horizontal scattering plane. When necessary, goniometers were used to reach out-of-plane momentum position. It should be stressed that MA and FA molecules were not deuterated giving rise to a large incoherent neutron scattering from the various hydrogen atoms from the organic part of the compounds. The fact that the samples are fixed onto a vanadium sheet also results in additional incoherent elastic scattering.\\ \\
\noindent \textbf{Raman Scattering Spectroscopy} 

Raman scattering in MAPB has been performed under an optical microscope and a T64000 Jobin-Yvon double pass diffractometer working with 18000\,trts/mm gratings. The radiation of a krypton laser emitting at 647\,nm was tightly focused into the sample with a x100 objective. In order to avoid photo-induced effects, the incident power was always kept lower than 2\,mW. The spectra have been obtained in the backscattering geometry with the incident light parallel to the [001]-cubic crystallographic direction of the sample. The incident polarization was parallel to the [110] direction and the results shown in Figures 2 and 3 of the main text correspond to the polarized spectra (\rm{i.e.} scattered light // [1${\bar 1}$0]). \\ \\
\noindent \textbf{Experimental data fitting.} The experimental TAS  spectra can be described by a sum of phonon terms on top of a flat background (BG). Each phonon is  typically accounted for by a damped harmonic oscillator (DHO) \cite{ATCNB}. The neutron spectra  can then be described by the following expression: 
\begin{equation}
\resizebox{0.9\hsize}{!}{$I({\bf Q},\omega) = BG + \Big[1-exp(-\frac{\hbar \omega}{k_b T})\Big]^{-1} \\ 
 \sum _{j} {|F_j({\bf Q})|^2  \frac{\omega \Gamma_j}{(\omega^2-{\omega_j}^2)^2 +(\omega \Gamma_j)^2}} $}
\label{DHO}
\end{equation}
where $\omega_j$ represents the energy, $\Gamma_j$ the damping and $F_j({\bf Q})$ is the dynamical structure factor of the j-th phonon. The number of mode depends on how many peaks are visible.  The prefactor of DHOs is the phonon population factor. This full scattering function, $I({\bf Q},\omega)$,  is next convoluted by the 4D spectrometer resolution function of the instrument as explained in \cite{ATCNB}, and used to fit the experimental data. The presence of several phonon lines in a given phonon bundle would have the effect to increase the damping when fitted by a DHO due to the energy resolution function. \\

\noindent \textbf{Acknowledgements.} This project has received funding from the European Union’s Horizon 2020 programme, through a FET Open research and innovation action under the grant agreement No 687008. The open access fee was covered by FILL2030, a European Union project within the  European  Commission’s Horizon  2020  Research  and Innovation programme under grant agreement N°731096. J. E. is senior member of institut universitaire de France.  We wish to thank Marc de Boissieu, Stéphane Pailhès and Yvan Sidis for stimulating discussions and Anna 
Bourges-C\'elaries for a proofreading of the manuscript. \\ 
\noindent \textbf{Author contributions.} P.B., C.K. and J.E. conceived and supervised the project; A.C.F. and P.B. performed the INS experiments at LLB Saclay and ILL Grenoble with support from A.L.; S.R. and J.O. participated in the INS experiments at ILL as the local contacts for IN12 and IN5, respectively; A.C.F. and P.B. analysed the neutron data; B.H. and B.R. performed the Raman scattering experiments and analysed the Raman data; S.P. synthesized the perovskite single crystal samples with support from S.C; J.E. contributed to the analysis and discussion of the results. A.C.F., P.B. and J.E. wrote the manuscript with further contributions from all authors. All authors contributed to this work, read the manuscript and agree to its contents. \\ \\
\noindent \textbf{Competing Interests.} 
The authors declare no competing interests.\\ 
\noindent \textbf{Data Availability}. Data collected on IN12 and IN5 are available at https://doi.ill.fr/10.5291/ILL-DATA.7-02-172 and https://doi.ill.fr/10.5291/ILL-DATA.TEST-2912, respectively. The rest of the data that support the findings of this study is available from the corresponding authors upon request. \\ \\
%
%

%
%
} 



\clearpage

\section{Supplementary Material}
 \beginsupplement

\onecolumngrid 

\section{\textbf{Supplementary figure 1: Room temperature Time-of-flight  neutron spectra}}
\begin{figure}[h!]
\centering
\includegraphics[width=0.65\linewidth]{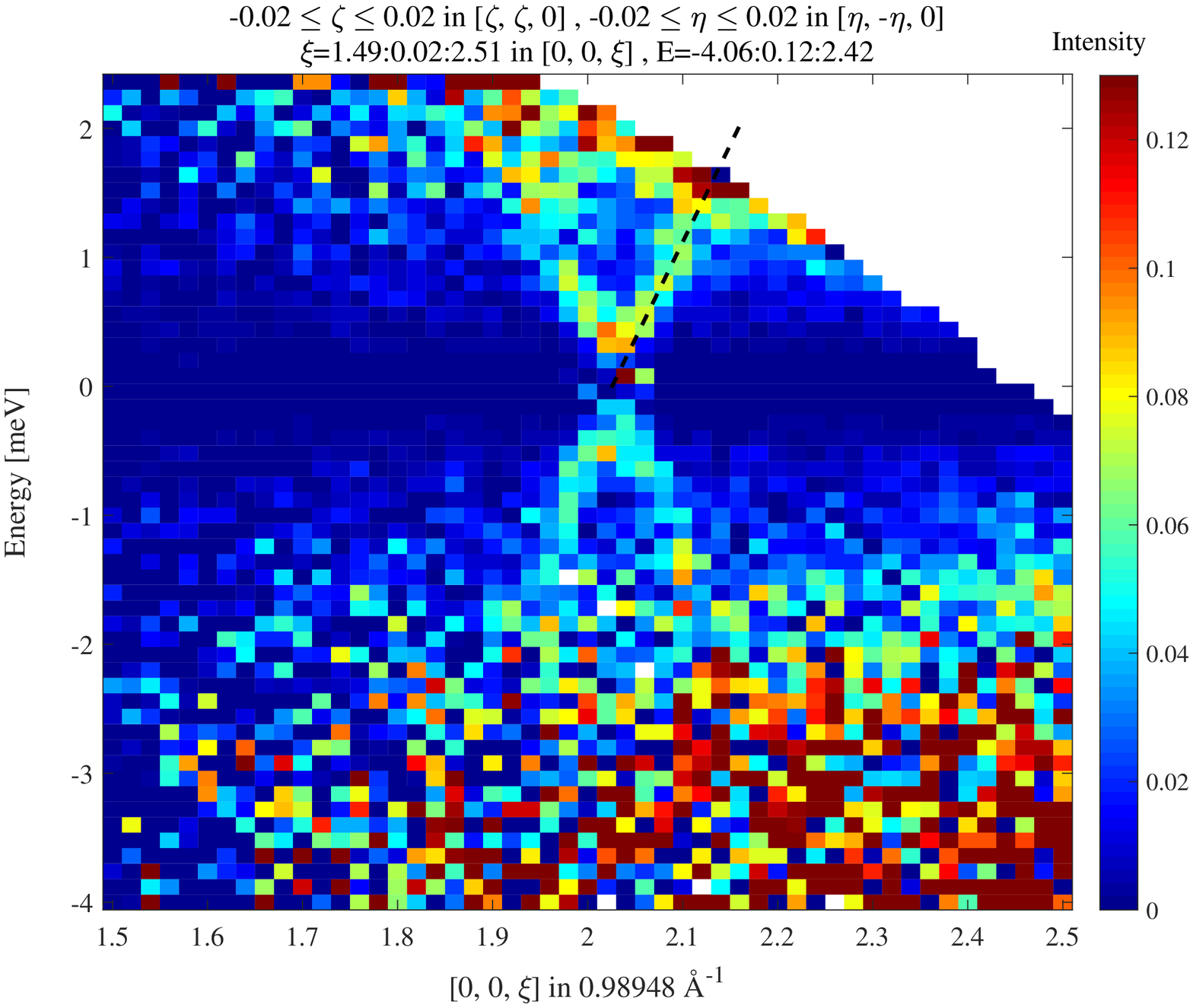}
\caption{
\textbf{Room temperature intensity map}. Time-of-flight (TOF) neutron spectra  measured at 300K in MAPbI$_3$. Longitudinal acoustic (LA) phonons around the 002 Bragg reflection are clearly seen up to energies of $\sim$2~meV. The slope (dashed line), corresponding to the sound velocity from the  previously reported dispersion curves of the same LA phonons \cite{Ferreira2018}, is also plotted for comparison.
}
\label{fig:LA-004}
\end{figure}
%

\section{\textbf{Supplementary figure 2: Room temperature neutron spectra}}
\begin{figure}[h!]
\centering
\includegraphics[width=0.6\linewidth]{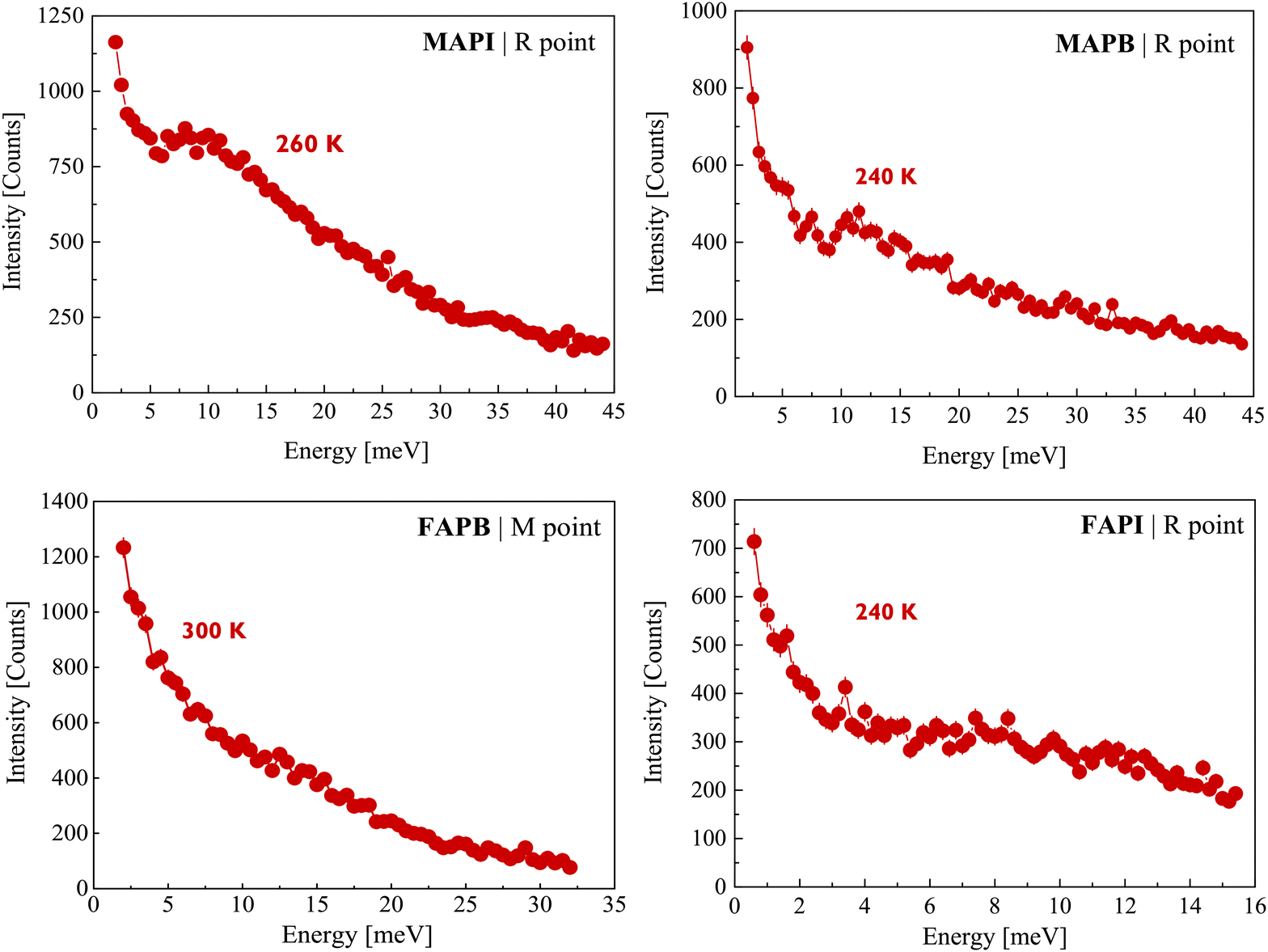}
\caption{
\textbf{High temperature inelastic neutron spectra spectra.} Optical phonon spectra, measured by  triple-axis spectrometer (TAS) inelastic neutron scattering, in MAPbI$_3$, MAPbBr$_3$, FAPbBr$_3$ and $\alpha$-FAPbI$_3$. Despite the different temperatures, all four scans show the absence of clear phonon modes as a result of them being overdamped.
Error bars represent one standard deviation.}
\label{fig:OptPhon-RT}
\end{figure}

\cleardoublepage

\section{\textbf{Supplementary figure 3: Low temperature phonon dispersion in FAPB}}
\begin{figure}[h!]
\centering
\includegraphics[width=0.7\linewidth]{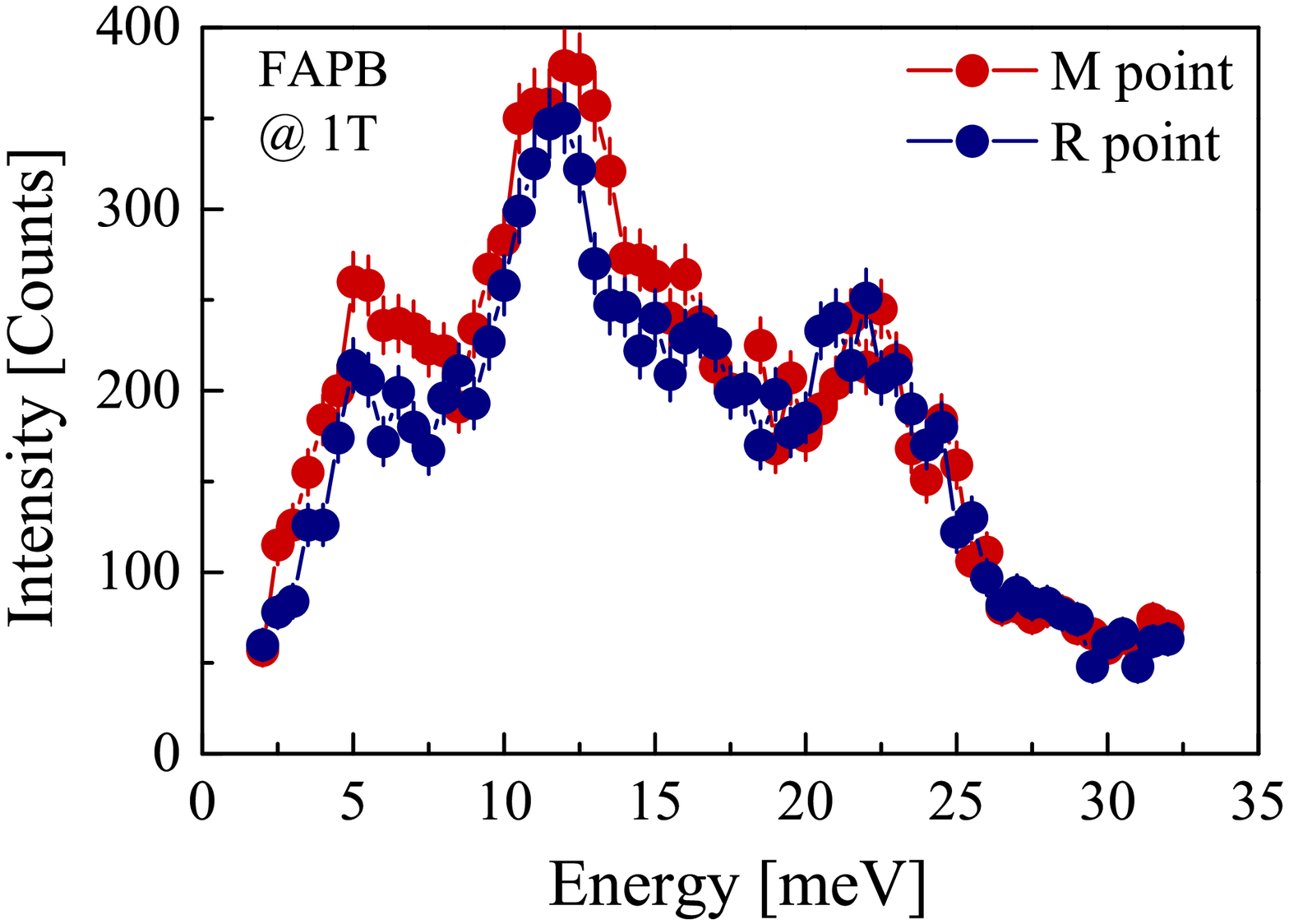}
\caption{
\textbf{Optical phonon dispersion in FAPB.} Optical phonon spectra, measured by triple-axis spectrometer (TAS) inelastic neutron scattering at low temperature (5~K), in FAPbBr$_3$ at the R and M Bragg points. Only a small negligible difference (in amplitude) is seen between the spectra of both points, illustrating the weak dispersion nature of the phonon modes. Error bars represent one standard deviation.
}
\label{fig:FA-disp-supp}
\end{figure}
%
%
\section{\textbf{Supplementary Note 1: Coherent and incoherent neutron cross sections}} 
In both  triple-axis spectrometer and time-of-flight inelastic neutron scattering experiments, the measured neutron intensity is the convolution product of the neutron cross-section 
${{d^2\sigma}\over{d\Omega dE_f}}$ by the resolution function of the instrument  $R({\bf Q},\omega)$ \cite{Dorner}. The neutron cross-section correspond to scattered neutrons with final energy between $E_f$ and $E_f+dE_f$ and for a beam over a solid angle  
$d\Omega$. Both are defining a scattered volume in momentum and energy spaces as $d^3{\bf Q} d\omega$.  The measured neutron intensity at a  momentum ${\bf Q_0}$ and a neutron energy transfer  $\omega_0$ can then be written as,
\begin{equation}
I({\bf Q_0},\omega_0)= \int d\omega d^3{\bf Q} \   \Big({{d^2\sigma}\over{d\Omega d\omega}} \Big)  R({\bf Q-Q_0},\omega-\omega_0)
\label{general}
\end{equation}
\noindent 
where $R(Q-Q_0,\omega-\omega_0)$ is the resolution function of the  triple-axis spectrometer or the time-of-flight instrument (note that the $k_F/k_I$ factor is included in the resolution function \cite{Dorner}). The neutron cross section is the sum of both coherent and incoherent cross sections~\cite{Squires,Lovesey}.
\begin{equation}
 \Big( {{d^2\sigma}\over{d\Omega d\omega}} \Big) =  \Big({{d^2\sigma}\over{d\Omega d\omega}} \Big)_{coh} +   \Big({{d^2\sigma}\over{d\Omega d\omega}} \Big)_{inc}
\label{coh-incoh}
\end{equation}
The coherent cross section corresponds to the correlations of atomic displacements of all nuclei at different times whereas the incoherent  cross section represents self-correlations only at different times of the same nucleus. The latter includes interference effect which are absent in the incoherent scattering.  \\
\indent All nuclei are characterized by two different  neutron scattering lengths, $b_{coh}$ and $b_{inc}$. The value of neutron scattering lengths for each nucleus and its isotope can be found on the website of the National Institute of Standards and Technology (NIST) center for  neutron research (https://www.nist.gov/ncnr/planning-your-experiment/sld-periodic-table). The value of $b_{coh}$ of each nuclei present in the hybrid organolead perovskites (HOP) is ranging between 9.4~$fm$ for Pb, to 5.28~$fm$ for I. The incoherent scattering length of hydrogen, $b_{inc}=25.27 fm$, is much larger than the ones of all other atoms (Pb, I, Br, C ,N) present here ($b_{inc}\le 2 fm$).  Therefore, as it is well-known for organic compounds, the incoherent cross section is mostly controlled by the hydrogen contribution. We also remind that the methylammonium (MA), CH$_3$NH$_3$, and formamidinium (FA), (CH$_2$)$_2$NH, molecules in our samples were protonated, giving rise to a large incoherent neutron scattering from the 6 hydrogen atoms in MA (or 5 in FA), per formula unit of the HPO compounds. \\
\indent Both neutron cross-section can be written within harmonic lattice vibrations approach \cite{Squires,Lovesey}. That defines phonons having a dispersion relation $\omega_{j}({\bf q})$  (eigenvalue of the dynamical matrix), where ${\bf Q} = {\bf q} + {\bf \tau}$ and ${\bf \tau}$ is the momentum of a Bragg peak of the nuclear structure.\\
\indent The coherent cross section of phonon scattering for a non-bravais lattice is~\cite{Squires,Lovesey},
\begin{equation}
 \Big({{d^2\sigma}\over{d\Omega d\omega}}\Big)_{coh}=\frac{(2 \pi)^3}{V_0} \sum_{j{\bf q}} \delta({\bf Q-q-\tau)} \left|  \sum_d \{{\bf Q} \cdot {\bf e}^j_{d}({\bf q})\}  \frac{b^d_{coh}}{\sqrt{M_d}} \exp^{(-i {\bf Q}\cdot {\bf d})} \exp^{-W_d({\bf Q})} \right|^2 S_j ({\bf Q},\omega)\\
\label{sqwcoh}
\end{equation}
One sees that the cross section can be separated in two terms: a structure factor and an energy dependent spectral weight function, $S_j({\bf Q},\omega)$. In the structure factor,  $b^d_{coh}$, $M_d$ and ${\bf e}_{d}$ are respectively the position of the d-th atom in the unit cell,  the scattering length (in $fm$), the molar atomic mass (in $g$) and the polarization vector for the j-th phonon of the atom labelled $d$ within the unit cell. $W_d({\bf Q})$ is the Debye-Waller factor of $d$-th atom in the unit cell. $V_0$ is the volume of the unit cell of the crystal. \\
\indent In case of no phonon damping, the spectral weight function  of the j-th phonon is,
\begin{equation}
S_j({\bf Q},\omega)= \frac{1}{2 \omega_{j}}[(1+n(\omega_{j}))
\delta(\omega -\omega_{j})+n(\omega_{j})\delta(\omega +\omega_{j})]
\label{sqw}
\end{equation}
$n(\omega_{j})$ is the Bose factor. $\omega_{j}({\bf q})$ is the phonon energy which can disperse, its ${\bf q}$-dependence is omitted in Eq.~\ref{sqw} for simplicity. The first part in Eq. \ref{sqw} corresponds to phonon creation and the second part corresponds to phonon annihilation. When considering a non-vanishing damping constant, $\Gamma_j$, the phonon spectrum is described by a damped harmonic oscillator (DHO), and  equation (\ref{sqw}) becomes \cite{Lovesey}
\begin{equation}
S_j({\bf Q},\omega)= \{1+n(\omega)\} \;
\frac{\omega\Gamma_j}{(\omega^2-\omega_{j}^2)^2+\omega^2\Gamma_j^2}
\label{dho}
\end{equation}
which included  the temperature factor detailed balance $\{1+n(\omega)\}={1\over{1-exp(-{{\hbar\omega}\over{k_BT}})}}$. The DHO model (Eq.~\ref{dho}) for the phonon response is used to fit our neutron spectra (see the main text). \\
\indent Within the same harmonic approximation and using the same notations, the incoherent cross section  for phonon scattering is  written  \cite{Squires,Lovesey}.
\begin{equation}
 \Big({{d^2\sigma}\over{d\Omega d\omega}}\Big)_{inc}=  \sum_d  (b^d_{inc})^2
 \frac{1}{M_d} \exp^{-2W_d({\bf Q})}   \sum_{j{\bf q}}  | {\bf Q} \cdot {\bf e}^j_{d}({\bf q})|^2  S_j ({\bf Q},\omega)\\
\label{sqwinc}
\end{equation}
The last part of Eq.~\ref{sqwinc} corresponds to a phonon density of states (sum in momentum space of phonon modes).
In contrast to the coherent cross section, the specific ${\bf q}$ dependence of the j-th atom is lost. In principle, no information can therefore  be obtained for the phonon dispersion in Eq.~\ref{sqwinc}. Typically, the incoherent cross section has the shape of a broad continuum if the optical phonons disperse (examples of such situations are multiple in the literature~\cite{Lovesey}). However, it exhibits sharp features  in case of dispersionless optical phonons. We are facing this last situation in hybride perovskite, especially for MA-based compounds where the low temperature phonon spectra show energy resolution-limited peaks. \\
\indent From the large $b_{inc}^H$ of hydrogen atoms, one sees that the neutron spectra of phonons involving hydrogen correspond to the incoherent cross-section whereas all other atoms in HOPs contribute to the coherent cross section. 


\section{\textbf{Supplementary Note 2: Atomic mass and phonon energy}}
In Figure 6 of the main text, we have compared the energy of the bundles \textit{a}, \textit{b} and \textit{c} between each of the four compounds. In the harmonic approximation, the phonon energy is proportional to the square root of the atomic mass, $M$, of the halide atom/organic molecule involved in that vibration\cite{Lovesey}. Therefore, one expects on general grounds, 
\begin{equation}
\omega_j \propto \frac{1}{\sqrt{M}}
\end{equation} 
The atomic masses of each nucleus or molecule can be easily estimated: $M_{FA}$~=~45.1~g/mol; $M_{MA}$~=~32.1~g/mol; $M_{Pb}$~=~205.0~g/mol; $M_I$~=~126.9~g/mol; $M_{Br}$~=~79.9~g/mol. One can then estimate relative energies of a given phonon when changing the molecule MA to FA or the halide I to Br when assuming the same atomic interactions.
So, for a phonon where only the halide atom is involved, one expects: 
\begin{equation}
\frac{\omega_j(iodines)}{\omega_j(bromides)} = \sqrt{\frac{M_{Br}}{M_I}} = 0.79.
\end{equation} 
In case the lead atom also participates to the vibration, one expects instead: 
\begin{equation}
\frac{\omega_j(iodines)}{\omega_j(bromides)} = \sqrt{\frac{M_{Pb}+3 M_{Br}}{M_{Pb}+3  M_I}} = 0.87.
\end{equation} 
\noindent In both case, the phonon energy of iodines will be lower than for bromides. In Fig. 5 of the main text, this corresponds to
the bundle \textit{a}, where one can estimate: $\frac{\omega_a(MAPI)}{\omega_a(MAPB)}$=0.6 and $\frac{\omega_a(FAPI)}{\omega_a(FAPB)}$=0.7. 
One notices that the measured effect is even larger than the estimated one. That suggests that the interactions responsible for that phonon bundles are weaker for bromides than iodines as the phonon energy is $\omega_j \propto \sqrt{ \frac{k}{M}}$ where $k$ represents atomic forces. 
For a given phonon where only the organic molecule is involved, one expects: 
\begin{equation}
\frac{\omega_j(FA)}{\omega_j(MA)} = \sqrt{\frac{M_{MA}}{M_{FA}}}=0.84.
\end{equation} 
The phonon energy is lower for the heavier FA molecule compared to MA. That corresponds to the bundle  \textit{c} where one can estimate
$\frac{\omega_c(FAPB)}{\omega_c(MAPB)}$=0.96 and $\frac{\omega_c(FAPI)}{\omega_c(MAPI)}$=0.64. Here again, one sees that the experimental trend does not exactly match the prediction from the above relationship from atomic masses. \\

\vspace{1cm}
\section{\textbf{Supplementary Note 3: Sample Preparation}} 
\textit{Methylammonium Lead Bromide}: MAPbBr$_3$ single crystals were grown by inverse temperature crystallization (ITC) \cite{saidaminov2015high}. A solution of MAPbBr$_3$ was prepared in N,N-dimethylformamide (DMF) solvent with 1 M concentration and was filtered with a 0.2~$\mu$m pore size polytetrafluoroethylene (PTFE) filters. 3~ml of the obtained solution were then placed into a 5~ml beaker which was introduced in an oven at 80~\degree C and kept for 3~h. To increase their size, the formed crystals were extracted from the first beaker and place into another beaker containing fresh filtered solution at the same temperature overnight.\\
\textit{Methylammonium Lead Iodide}: MAPbI$_3$ single crystals were grown by ITC \cite{zhumekenov2016formamidinium}. A solution with 1~M concentration of MAPbI$_3$ was prepared in $\gamma$-butyrolactone (GBL)  solvent and was filtered with a 0.2~$\mu$m pore size PTFE filters. Then 3~ml of the obtained solution were placed into a 5\,ml vial which was placed in an oven at 60~\degree C. The temperature was gradually increased to 110~\degree C and kept for 1 days to further increase the size of the crystals.\\
\textit{Formamidinium Lead Bromide}: FAPbBr$_3$ single crystals were grown by ITC \cite{saidaminov2015high}. After the filtration using PTFE filters with a 0.2\,$\mu$m pore size, 3~ml of 1~M solution of FAPbBr$_3$ in DMF:GBL (1:1 v/v) were placed into a 5~ml beaker which was introduced in an oven at 40~\degree\,C. The temperature was then gradually increased to 52~\degree C and kept for 5~h and at 60~\degree C for 3~h. The size of the crystal can be further increased through the gradual increase of temperature.\\
\textit{Formamidinium Lead Iodide}: FAPbI$_3$ single crystals were grown by ITC \cite{zhumekenov2016formamidinium}. A solution of FAPbI$_3$ was prepared in GBL with 1\,M concentration and was filtered with a 0.2\,$\mu$m pore size PTFE filter. Then 3~ml of the obtained solution were placed into a 5~ml vial which was immersed in an oil bath at 80~\degree C. The temperature was slowly increased to 105~\degree C. Subsequently a fresh filtered solution can be added on one formed crystal in a vial to increase the size through the gradual increase of temperature. \\
\textit{Note:} The $\alpha$-phase of FAPbI$_3$, \textit{i.e.} the photoactive phase, is metastable and only lasts a maximum of 7 days at room temperature. In this work we were able to measure optical phonons on a fresh sample within that first period of 7 days, mainly due to the low temperature (5~K) working conditions. Only once we started heating up the sample, for the temperature dependent measurements, it did start showing signs of degradation into the yellow phase ($\delta$-phase). The $\alpha$-phase can, however, be restored back into the black $\alpha$-phase upon heating/annealing, which was indeed demonstrated in small single crystals \cite{zhumekenov2016formamidinium}. However, as we mentioned in our previous study on acoustic phonons \cite{Ferreira2018}, on large single crystals such as ours, only part of the sample is restored to a single grain and most of the sample remains as a powder.\\ \\
%

%
%

\section{\textbf{References:}}

\bibliography{BibOptPhon}

\end{document}